\documentclass[final,5p,times, a4paper]{elsarticle}

\usepackage{amssymb}
\usepackage{hyperref}

\journal{Science of Computer Programming}

\begin{document}

\begin{frontmatter}




\title{An Event-Driven Tool for Context-Aware Code Smell Detection Using SmellDSL}


\author[label1]{Matheus dos Santos Viegas}
\ead{viegasmatheus@edu.unisinos.br}
\author[label1]{Adrian Gabriel Keller dos Santos}
\ead{santosadrian@edu.unisinos.br}
\author[label1]{Kleinner Farias}
\ead{kleinnerfarias@unisinos.br}
\author[label1]{Robson Keemps da Silva}
\ead{robson.keemps@edu.unisinos.br}

\address[label1]{Applied Computing Graduate Program, University of Vale do Rio dos Sinos (Unisinos), São Leopoldo, RS, Brazil }

\begin{abstract}
Code smells signal violations of design principles that degrade the internal quality of evolving software systems. Although many tools detect such anomalies using static metrics, they often ignore the development context in which smells arise and are resolved. This limitation can lead to misleading warnings and weak support for refactoring decisions. To address this problem, we present \texttt{SmellHunter}, a context-aware tool that interprets scripts written in the SmellDSL domain-specific language to detect and contextualize code smells. \texttt{SmellHunter} integrates static code metrics with contextual information (such as team characteristics, project stage, and geographic metadata) to produce richer, more actionable analyses. The tool adopts an event-driven architecture in which a service bus orchestrates validation, interpretation, and persistence services through asynchronous events. This architecture enables scalable analysis while minimizing disruption to developers' workflows. \texttt{SmellHunter} is integrated into the Eclipse development environment via a dedicated plugin and provides aggregated insights via a mobile application, allowing developers to explore smell occurrences by type, severity, and location. By linking smell detection with contextual data and collaborative visualization, \texttt{SmellHunter} supports developers acting as \textit{smell hunters}, helping teams identify recurring quality issues emerging from a particular location and assign refactoring tasks to developers with relevant expertise. We describe the architecture of \texttt{SmellHunter}, the interpretation process of SmellDSL scripts, and the integration of contextual data to support more informed refactoring decisions in modern software development environments.
\end{abstract}


\end{frontmatter}


\section*{Metadata}
\label{lab:metadata}

The metadata for the proposed tool are shown in Table \ref{tab:metadata}.


\begin{table*}[!ht]
\centering
\begin{tabular}{|l|p{4.5cm}|p{6.5cm}|}
\hline
\textbf{Nr.} & \textbf{Code metadata description} & \textbf{Please fill in this column} \\
\hline
C1 & Current code version & V1.0 \\
\hline
C2 & Permanent link to code/repository used for this code version &
\url{https://github.com/MathDEV-0/SmellHunterAPI} \\
& & \url{https://github.com/MathDEV-0/SmellHunter-Eclipse-Plugin} \\
\hline
C3 & Permanent link to Reproducible Capsule &  \\
\hline
C4 & Legal Code License & MIT License \\
\hline
C5 & Code versioning system used & Git \\
\hline
C6 & Software code languages, tools, and services used &
Python, Flask (interpreter); \newline
Java, Eclipse RCP (plugin) \\
\hline
C7 & Compilation requirements, operating environments and dependencies &
Python 3.10+; \newline
Java 21+, Eclipse RCP \\
\hline
C8 & If available, link to developer documentation/manual &
SmellHunter API: \url{https://github.com/MathDEV-0/SmellHunterAPI/blob/main/README.md} \\
& &
SmellHunter Eclipse Plugin: \url{https://github.com/MathDEV-0/SmellHunter-Eclipse-Plugin/blob/main/README.md} \\
& & Demonstration: \url{https://www.youtube.com/watch?v=WKONlb5o1TY} \\
\hline
C9 & Support email for questions & matheusdsantviegas@gmail.com \\
\hline
\end{tabular}
\caption{Code metadata.}
\label{tab:metadata}
\end{table*}

\section{Motivation and significance}

Code smells are symptoms of poor design decisions in source code and can impair software maintainability \cite{nocera2026icse,oizumi2016code}. Over time, these smells degrade design quality, increase maintenance costs, and may compromise architectural stability~\cite{fowler2018refactoring,diaz2025global,martins2021cooccurrences}. Moreover, code smells can significantly increase change-proneness \cite{nocera2026icse}. Code smells are therefore widely regarded as early indicators of design problems that require refactoring \cite{tsantalis2022refactoringminer} to prevent further deterioration of the software structure~\cite{fowler2018refactoring}. Numerous tools have been proposed to support smell detection, including SonarQube\footnote{SonarQube: \url{https://www.sonarsource.com/products/sonarqube/}}, Qodana\footnote{Qodana: \url{https://www.jetbrains.com/qodana/}}, and RefactoringMiner~\cite{hilmi2023research,tsantalis2022refactoringminer}. While these tools provide robust static analysis capabilities, refactoring heuristics, and rely on predefined metrics and rules, they often treat smells as universal violations of coding conventions. In practice, however, not all smells carry the same weight for every team \cite{keemps2025smelldsl,hozano2018you}. Development teams often perceive certain smells as more critical \cite{hozano2018you} because they threaten architectural stability or violate local coding practices shaped by project constraints and organizational culture \cite{keemps2025smelldsl}. More critically, prior studies \cite{oliveira2022developers,hozano2018you} reveal a marked lack of consensus among developers on how to detect even well-established code smells \cite{suryanarayana2014refactoring}, underscoring the inherently subjective nature of this process.

\texttt{SmellHunter} addresses this concern by supporting the detection of \textit{team-sensitive smells}. \texttt{SmellHunter} allows teams to specify priorities through SmellDSL scripts \cite{keemps2025smelldsl} and associate smell occurrences with contextual attributes, including project stage and geolocation. This functionality enables developers to identify where smells emerge and are resolved, allowing teams to map quality issues to specific regions, organizations, or development environments and to reveal locations where targeted improvements may be required. Prior empirical studies \cite{d2020effects,d2020sw} have demonstrated the importance of incorporating contextual information in software development practices. This work addresses a scientific and technical problem: the lack of tools to interpret executable specifications of smell-detection rules, expressed in SmellDSL, in a context-aware development environment. SmellDSL is a domain-specific language to specific bad smells \cite{keemps2025smelldsl}.

Existing approaches allow developers to define smell detection rules, but they rarely provide mechanisms to dynamically interpret these rules or integrate them with contextual signals that influence development decisions~\cite{albuquerque2023integrating,keemps2025smelldsl}. As a result, many existing tools remain limited to static rule evaluation and often fail to capture the broader context in which smells emerge. \texttt{SmellHunter} closes this gap by interpreting scripts written in the \textit{SmellDSL} language and generating automated analyses of code anomalies enriched with contextual metadata. Developers receive feedback directly within the Eclipse development environment and through a complementary mobile application. This integration transforms smell detection from a static rule-based activity into an interactive, operational process. The scientific significance lies in enabling executable smell specifications that can be validated, interpreted, and acted through automated analysis pipelines that align technical metrics with the real constraints and priorities of development teams.

Our work also opens new opportunities for scientific investigation in context-aware code quality analysis. Traditional smell detection mechanisms typically operate in batch, analyzing code snapshots with static metrics~\cite{hilmi2023research}. Such approaches rarely capture the dynamic and contextual nature of software development activities. \texttt{SmellHunter} reframes smell detection as \textit{an asynchronous, event-driven process} that integrates contextual information into the analysis pipeline. Each smell-detection request becomes an event processed by a service bus that coordinates validation, interpretation, and persistence services. Our proposed architecture (Section \ref{sec:architectire}) enables large-scale analysis while preserving the responsiveness and traceability of smell detection activities. Furthermore, by associating smell occurrences with contextual and geographical metadata, \texttt{SmellHunter} introduces the concept of \textit{location-aware quality analysis}. Developers may specialize in detecting and resolving specific smell categories and act as ``smell hunters'' responsible for particular contexts or regions. Researchers can therefore investigate spatial and organizational patterns of code degradation and study how team practices influence the emergence and resolution of smells.

From an operational perspective, our work supports development teams engaged in continuous software evolution. Teams define smell specifications using SmellDSL \textit{script} and configure \textit{thresholds} for relevant metrics based on their development practices and \textit{quality priorities}. Through the Eclipse plugin, developers submit analysis requests that trigger asynchronous processing in the \texttt{SmellHunter} backend. A \textit{validation service} verifies the script syntax and threshold definitions, while the \textit{interpretation service} evaluates the rules against project artifacts. Detected smells are stored and disseminated through \textit{client interfaces}. Developers receive notifications in the IDE and through a mobile interface that summarizes smell occurrences by severity, type, and location. This workflow enables teams to detect and address code smells while maintaining software in parallel, reducing disruption to ongoing development and supporting continuous maintenance.

Considering the implementation of \texttt{SmellHunter}, it was built upon widely adopted technologies and frameworks that ensure scalability and interoperability with modern development infrastructures. The tool integrates with the Eclipse Platform, the primary development environment for executing SmellDSL scripts. The backend services are implemented in Python, leveraging the Lark parsing library to interpret the SmellDSL grammar and execute rule specifications. The service layer employs the Flask framework to expose analysis endpoints and orchestrate event-driven processing. Detected smell data and contextual metadata are stored in Google Cloud infrastructure, ensuring scalable persistence and distributed processing. The system also integrates with Google AppSheet to provide mobile visualization and interaction capabilities, enabling developers to inspect smell occurrences and access contextual information beyond the IDE. Together, these technologies form a robust platform that supports the distributed and asynchronous nature of context-aware smell detection.

\texttt{SmellHunter} differs from existing smell detection tools in three main aspects. First, it introduces an executable interpretation mechanism for SmellDSL scripts, enabling developers not only to define smell rules but also to execute and validate them automatically within an integrated environment. Secondly, our tool adopts an event-driven architecture that treats smell detection as an asynchronous pipeline, improving scalability and enabling decoupled processing through services responsible for validation, interpretation, and persistence. Third, \texttt{SmellHunter} incorporates contextual and geolocation information into smell analysis, allowing teams to identify where smells emerge and which developers or locations are most effective in resolving them. These features collectively extend traditional smell detection approaches and position \texttt{SmellHunter} as a platform for context-aware software quality monitoring, capable of aligning technical analysis with the collective practices and priorities of development teams.

\section{Software description}

\subsection{Software architecture}
\label{sec:architectire}

\begin{figure*}[!ht]
    \centering
    \includegraphics[width=\textwidth]{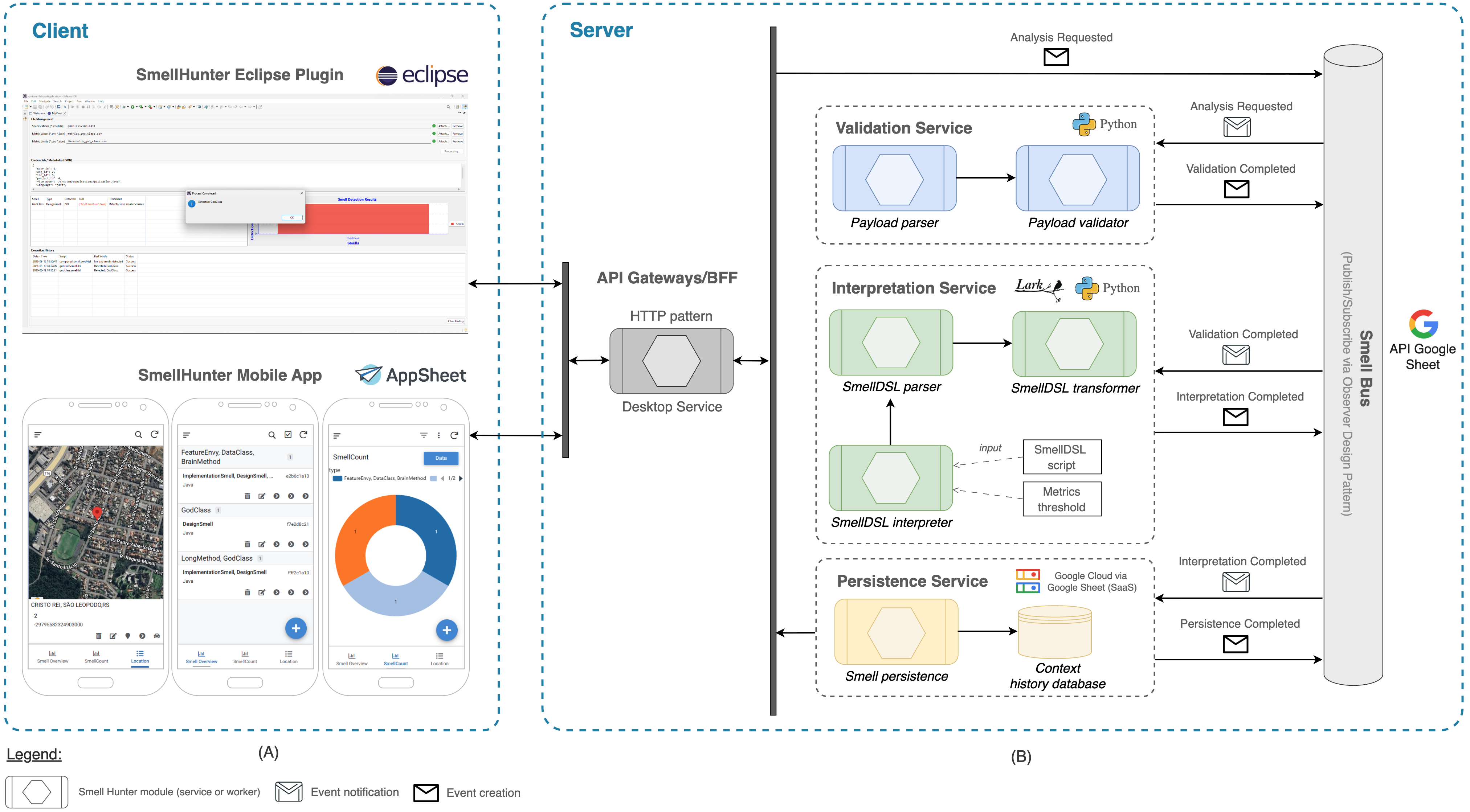}
    \caption{Event-driven architecture of \texttt{SmellHunter}. 
    (A) Client environment composed of the \texttt{SmellHunter} Eclipse plugin and the \texttt{SmellHunter} mobile application. 
    (B) Server environment containing validation, interpretation, and persistence services coordinated through the Smell Bus.}
    \label{fig:smellhunter_architecture}
\end{figure*}

Figure~\ref{fig:smellhunter_architecture} presents an overview of the \texttt{SmellHunter} architecture, which follows an event-driven architectural approach \cite{stopford2018designing}. Our system is organized into two main environments: the \textit{Client} and the \textit{Server}. This separation promotes modularity, scalability, and loose coupling among components while enabling asynchronous processing of smell detection requests.

The \textbf{Client} environment (Figure~\ref{fig:smellhunter_architecture}A) comprises two applications that interact with developers and smell hunters. The first is the \textit{SmellHunter Eclipse Plugin}, which integrates with the Eclipse development environment. This plugin allows developers to write and execute SmellDSL scripts, configure metric thresholds, and trigger smell analysis requests directly from the IDE. The second client component is the \textit{SmellHunter Mobile App}, implemented using Google AppSheet\footnote{Google AppSheet: https://about.appsheet.com/home/}, a low-code platform that connects our data and system on a single platform and is fully integrated with Google Workspace. The mobile application allows users to visualize detected smells, inspect their severity and type, and analyze their geographical distribution through geolocation features. 
This interface enables developers to explore quality issues across teams and locations, supporting the concept of ``\textit{smell hunters}'' who specialize in detecting and resolving specific anomalies.

The \textbf{Server} environment (Figure~\ref{fig:smellhunter_architecture}B) hosts the services that process smell detection requests. Communication between the client applications and backend services occurs through an \textit{API Gateway/BFF} (Backend for Frontend), which exposes an HTTP-based interface that mediates requests from the Eclipse plugin and mobile application. 
Once a request is received, the system publishes an \textit{Analysis Requested} event to the system’s communication backbone, the \textit{Smell Bus}. 
This service bus implements the \textit{Observer design pattern}, allowing independent services to subscribe to and react to events without direct dependencies.

The first service activated in the processing pipeline is the \textit{Validation Service}, implemented in Python. This service includes two main modules: a \textit{payload parser} and a \textit{payload validator}. It verifies the syntactic and semantic correctness of the analysis request, ensuring that the SmellDSL script, the associated metrics, and the defined thresholds are valid. 
When validation succeeds, the service emits a \textit{Validation Completed} event to the \textit{Smell Bus}. The \textit{Interpretation Service} then subscribes to this event and begins interpreting the validated SmellDSL script. This service uses the \textit{Lark}\footnote{Lark: \url{https://lark-parser.readthedocs.io}}, a modern parsing library for Python that can parse any context-free grammar. It processes the grammar of the SmellDSL language and transform script specifications into executable detection rules. 

The service includes a \textit{SmellDSL parser}, a \textit{SmellDSL transformer}, and a \textit{SmellDSL interpreter}. 
These modules evaluate the smell rules against the input data and determine whether a code anomaly is present. 
After the interpretation phase, the service publishes an \textit{Interpretation Completed} event containing the detection results.

Finally, the \textit{Persistence Service} consumes the interpretation events and stores smell detection data in a \textit{context history database} hosted in the Google Cloud environment. 
This service records contextual information such as smell type, severity, and geolocation metadata. 
Once persistence is completed, a \textit{Persistence Completed} event is generated and propagated through the Smell Bus. 
This event-driven workflow enables asynchronous processing and ensures that each service can evolve independently while remaining coordinated through the event infrastructure.

\subsection{Software functionalities}

\texttt{SmellHunter} provides a set of functionalities designed to support context-aware detection and analysis of code smells in distributed development environments:

\begin{enumerate}
 \item \textit{Smell specification using SmellDSL:} Developers can define code smell detection rules using the SmellDSL domain-specific language, including metric thresholds and anomaly patterns.

 \item \textit{Automated validation of smell definitions:} The system validates SmellDSL scripts and associated metrics before execution, ensuring that analysis requests are syntactically and semantically correct.

 \item \textit{Event-driven smell detection pipeline:} Smell detection requests are processed asynchronously through the Smell Bus, enabling scalable analysis and loose coupling between validation, interpretation, and persistence services.

 \item \textit{Context-aware smell monitoring:} Detected smells are enriched with contextual attributes such as severity level, development context, and geolocation metadata, enabling deeper analysis of software quality issues.

 \item \textit{Multi-platform visualization and monitoring:} Developers receive feedback through both the Eclipse plugin and the mobile application, which allows users to explore smell occurrences, identify hotspots, and monitor trends in smell generation and resolution.
\end{enumerate}

These functionalities collectively enable development teams to identify critical code anomalies, track their evolution over time, and prioritize refactoring efforts based on contextual information.

\subsection{Sample use case}

A typical use case begins with a developer defining a smell-detection rule using a SmellDSL script in the Eclipse plugin. The developer specifies the relevant metrics and thresholds that characterize a particular anomaly. Once the script is executed, the plugin sends an analysis request to the server through the API Gateway. The request generates an \textit{Analysis Requested} event that is published to the \textit{Smell Bus}. The \textit{Validation Service} consumes this event and verifies the payload's correctness. If the request is valid, the system emits a \textit{Validation Completed} event that activates the \textit{Interpretation Service}. This service interprets the SmellDSL script and evaluates the defined rules against the analyzed software artifacts.

If a code smell is detected, the result is published as an interpretation event and subsequently consumed by the \textit{Persistence Service}, which records the detection in the cloud-based database along with contextual metadata. The results become immediately available in the Eclipse plugin and in the \texttt{SmellHunter} mobile application (Figure \ref{fig:mobile-interface}). Through the mobile interface, developers can visualize detected smells by type, severity, and location. This capability enables teams to identify regions or development contexts where anomalies frequently occur and allows specialized developers—acting as \textit{smell hunters}—to focus on resolving specific categories of code smells.

\section{Illustrative Examples}

\begin{figure*}[!ht]
    \centering
    \includegraphics[width=0.97\textwidth]{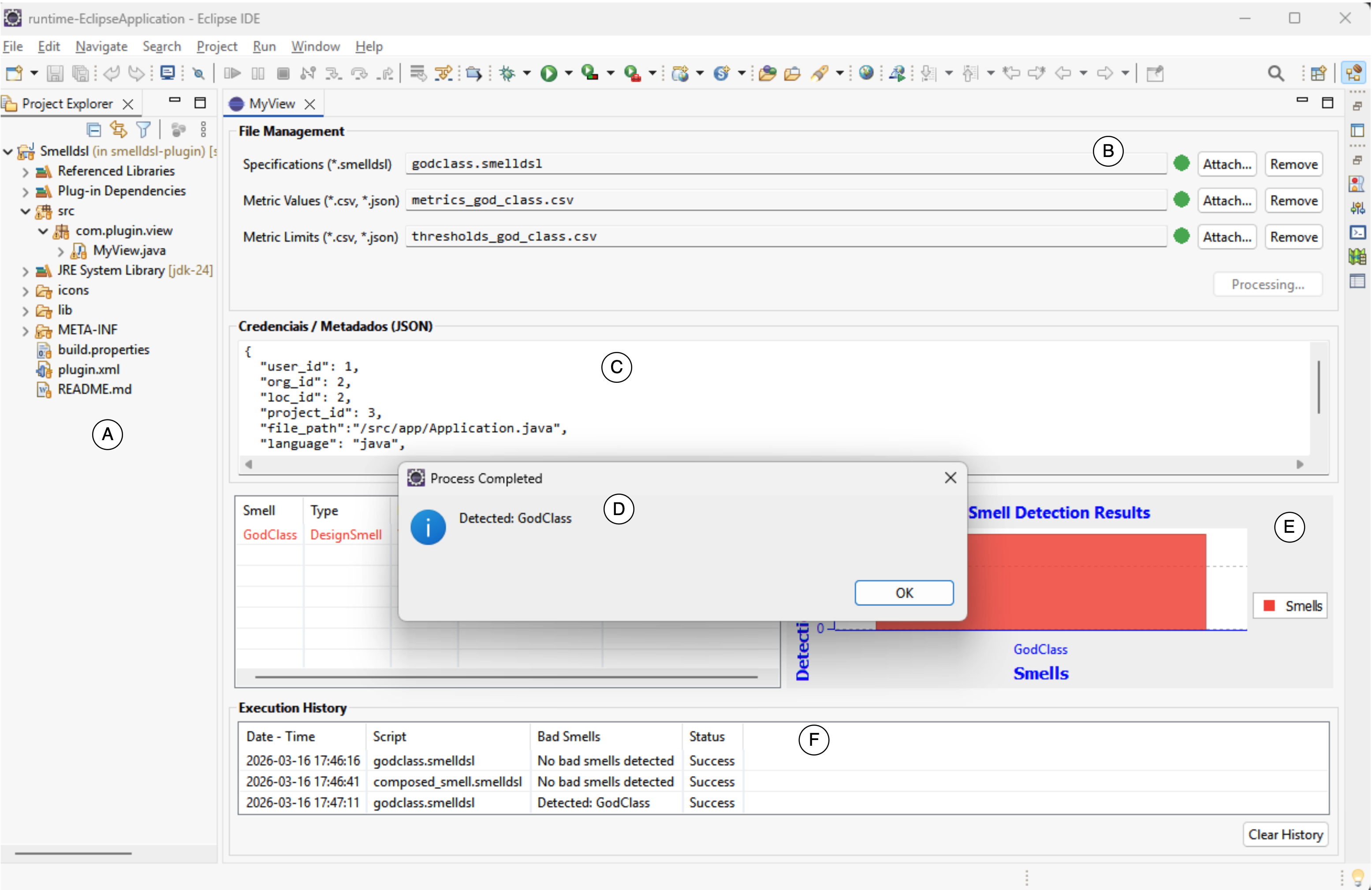}
    \caption{Illustrative example of the \texttt{SmellHunter} interface in the Eclipse environment. 
    (A) Project Explorer showing the structure of a SmellDSL project. 
    (B) Input specification panel with SmellDSL script, metrics, and thresholds. 
    (C) Metadata configuration in JSON format used in the analysis request. 
    (D) Execution feedback indicating the detection of a God Class smell. 
    (E) Histogram summarizing detected smells. 
    (F) Execution history displaying previous smell detection runs.}
    \label{fig:smellhunter_interface}
\end{figure*}

\textbf{Eclipse platform.}  Figure~\ref{fig:smellhunter_interface} illustrates an example of the \texttt{SmellHunter} tool integrated into the Eclipse development environment. 
The interface demonstrates how developers define smell detection specifications, submit analysis requests, and inspect the results produced by the system.

Part~(A) shows the \textit{Project Explorer}, where the developer manages the structure of a \textit{SmellDSL} project. In this view, users organize source files, plugin dependencies, and resources related to smell detection specifications. The project contains the SmellDSL scripts and the supporting configuration files required for the analysis process. This structure allows developers to maintain smell detection rules in the same environment used for software development.

Part~(B) highlights the \textit{input specification panel}. In this section, the user selects the files required for the smell detection request. These inputs include the SmellDSL specification file, the metric values file, and the metric threshold configuration file. The metric and threshold information is typically represented in CSV or JSON format, while the detection logic is defined in a SmellDSL script. Once these elements are attached, the tool prepares the payload that will be sent to the application server for analysis.

Part~(C) presents the \textit{metadata configuration panel}, where contextual information is defined in JSON format. This metadata includes attributes such as user identifier, organization identifier, project identifier, file path, and programming language. The information is transmitted with the smell detection request, enabling backend services to associate detected smells with their corresponding development contexts.

Part~(D) displays the \textit{execution feedback window}. After the server processes the request, the system returns a response indicating whether a smell has been detected. In the example shown, the analysis process completed successfully and identified the \textit{God Class} design smell. This immediate notification provides developers with quick feedback about potential anomalies in the analyzed code.

Part~(E) shows the \textit{Smell Detection Results} panel, which visualizes detected smells through a histogram. 
The chart summarizes the occurrences of smell types identified during the analysis. 
This graphical representation helps developers quickly understand the distribution of anomalies and prioritize refactoring efforts.

Finally, Part~(F) presents the \textit{execution history view}. This panel records the history of smell detection executions associated with specific SmellDSL scripts. Each entry includes the execution timestamp, the script used, the detection result, and the execution status. In the illustrated example, three executions were performed using different scripts, and one of them detected a \textit{God Class} smell. This historical view allows developers to track analyses over time and review previous detection outcomes.

\begin{figure*}[!ht]
    \centering
    \includegraphics[width=\textwidth]{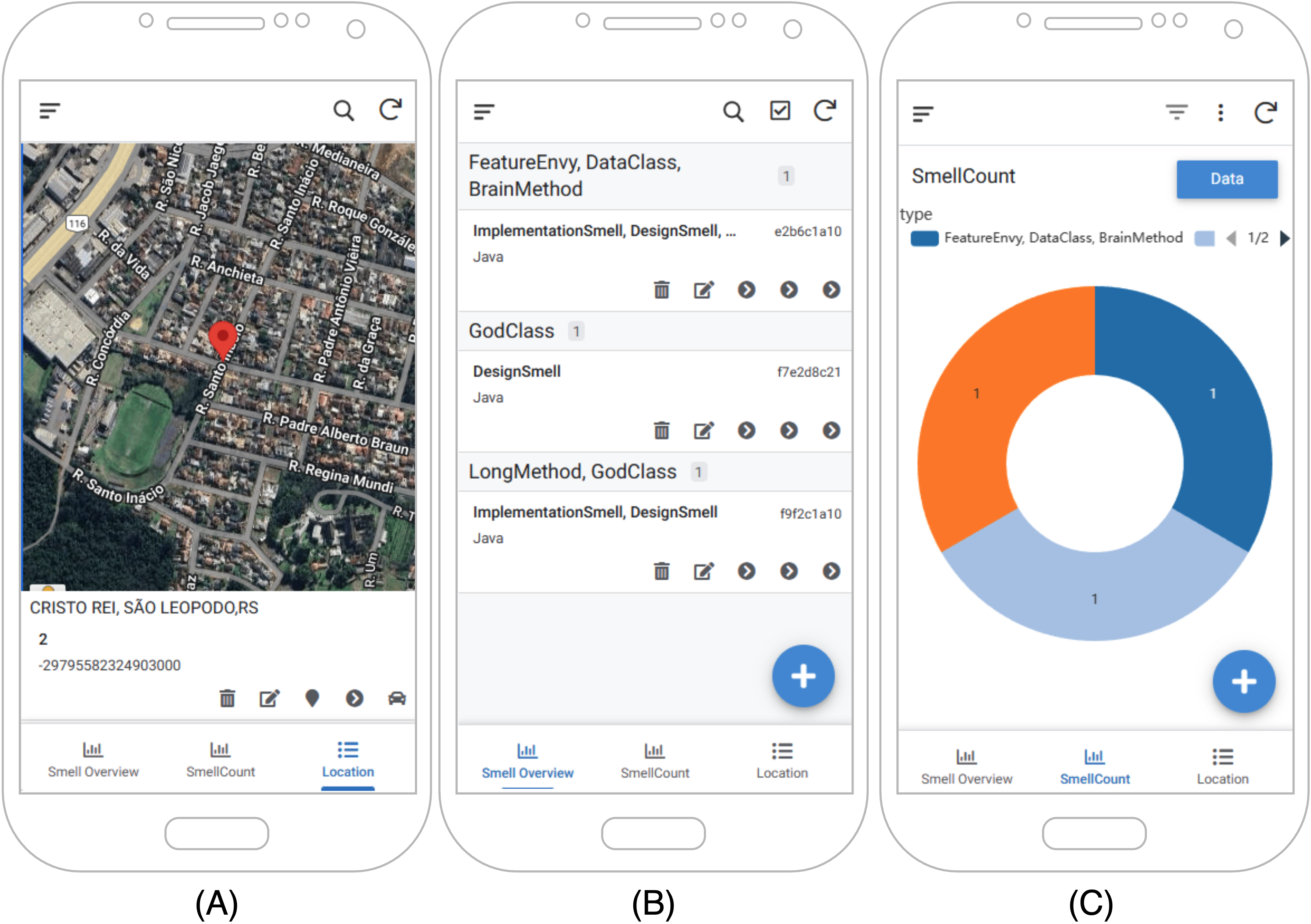}
    \caption{\texttt{SmellHunter} mobile application interface: (A) geolocation of detected bad smells, (B) list of detected smell types, and (C) dashboard summarizing smell occurrences.}
    \label{fig:mobile-interface}
\end{figure*}

Figure \ref{fig:mobile-interface} presents the \texttt{SmellHunter} mobile application interface. Part (A) shows the geolocation view, where the occurrence of a bad smell is associated with specific latitude and longitude coordinates, allowing users to identify where anomalies occur. Part (B) lists the detected types of bad smells, while Part (C) provides a dashboard that summarizes their distribution through a visual count, supporting quick analysis and prioritization.

\section{Impact} 

The \texttt{SmellHunter} tool contributes to ongoing efforts to better understand the role of context, subjectivity, and evolution in code quality assessment \cite{diaz2025global,hilmi2023research,pereira2022visualization}. Traditional smell detection approaches often focus on identifying anomalies from an individual developer’s perspective, relying primarily on static metrics and predefined thresholds. \texttt{SmellHunter} expands this view by supporting the detection of smells that are \textit{team-sensitive}, reflecting collective perceptions of quality degradation rather than individual preferences. 

By associating smell occurrences with contextual information—such as project identifiers, organizational context, and geolocation metadata—the tool raises the notion of what constitutes a bad smell to a collective level. 
This shift enables new research questions about how development teams define and prioritize quality issues, how perceptions of smell evolve across organizational contexts, and how contextual signals influence refactoring decisions. 
The tool also supports empirical investigations into the dynamics of smell emergence and resolution in distributed teams, providing a foundation for studying collaborative quality management in modern software ecosystems.

From a research perspective, \texttt{SmellHunter} strengthens the investigation of context-aware smell detection and asynchronous analysis architectures. 
Its event-driven infrastructure captures detailed information about smell detection requests, validation steps, interpretation outcomes, and persistence events. 
This data enables researchers to analyze the temporal and contextual evolution of smells across projects and teams. Such capabilities open opportunities for studying how developers specialize in resolving particular smell categories and how certain development environments or locations exhibit higher concentrations of anomalies. 
Moreover, by enabling automated interpretation of \textit{SmellDSL} specifications, the tool advances existing research questions related to domain-specific languages for software quality analysis, automated detection pipelines, and collaborative refactoring practices. These capabilities may also support future empirical studies exploring how contextual factors influence the interpretation of software metrics and the prioritization of design improvements.

In educational and industrial contexts, \texttt{SmellHunter} can influence daily development practices and training activities. In academic environments, the tool provides students with a practical platform for defining smell detection rules, interpreting detection results, and analyzing the evolution of design anomalies. This encourages reflective learning about design quality and collaborative code maintenance. In industrial settings, the integration of the Eclipse plugin and the mobile visualization interface enables development teams to continuously monitor bad-smell occurrences and coordinate refactoring efforts across distributed environments. 

By enabling teams to identify hotspots where smells frequently emerge and to track their resolution over time, the tool supports more systematic quality management processes. Furthermore, the event-driven architecture \cite{carbonera2020software} and cloud-based persistence infrastructure make the platform suitable for integration into industrial DevOps pipelines and large-scale software maintenance workflows.

\section{Conclusions}
\label{sec:conclusions}


This paper presented \texttt{SmellHunter}, a context-aware and event-driven tool designed to support the detection and interpretation of code smells specified through the \textit{SmellDSL} domain-specific language. The tool integrates with the Eclipse development environment through a dedicated plugin and complements this integration with a mobile application that visualizes detected smells and contextual information. Its architecture relies on asynchronous communication via the \textit{SmellBus}, which coordinates validation, interpretation, and persistence services to ensure modularity, scalability, and extensibility, while allowing developers to submit smell-detection requests and receive automated feedback without interrupting their workflow. 

Moreover, \texttt{SmellHunter} introduces a new perspective on smell detection by supporting \textit{team-sensitive smells}, in which the definition of what constitutes a bad smell emerges from collective decisions informed by SmellDSL specifications and contextual metadata, rather than from static rules alone. By associating smell occurrences with organizational, project, and location-based information, the tool elevates smell detection from an individual judgment to a collective and contextual understanding of software quality. Our results indicate that our tool can serve as a practical platform for maintaining and evolving continuously changing software systems by enabling developers to define smell detection rules, have them interpreted automatically, and visualize results in real time. In doing so, the tool strengthens development teams' ability to identify critical anomalies early, coordinate refactoring activities more effectively, and adopt more collaborative and context-aware practices in software quality management.

\section{Future Plans}

Future developments of \texttt{SmellHunter} will focus on improving its robustness, intelligence, and scalability. Key enhancements will include predictive capabilities that leverage historical and contextual data to anticipate code smells and assess their impact on maintainability, allowing earlier interventions and strategic refactoring. Another planned extension concerns automated knowledge synthesis and reporting. Architecturally, future versions will focus on increased scalability, fault tolerance, and responsibility isolation, adopting a modular, event-oriented design that supports independent analysis components and parallel processing. This will ensure responsiveness and reliability as project complexity increases. Plans include improving observability and traceability throughout the analysis pipeline to enable longitudinal studies of code quality and insights into refactoring effectiveness. Initially, efforts will focus on validating the analysis pipeline's correctness and integration with contextual data, laying the groundwork for advanced orchestration and intelligent recommendations. Overall, \texttt{SmellHunter} aims to provide enriched, decision-oriented support for code quality management.

Additionally, future work will extend \texttt{SmellHunter} toward more intelligent and adaptive software quality analysis environments. One promising direction is to incorporate cognitive and behavioral indicators into the smell detection process, enabling investigation of how developers perceive, interpret, and prioritize code anomalies under varying cognitive workloads~\cite{gonccales2021measuring}. Another direction concerns the use of contextual and historical smell data to support effort estimation and maintenance planning activities, helping teams better predict the impact of refactoring and software evolution tasks~\cite{carbonera2020software}. We also plan to investigate the scalability and modularity implications of the proposed event-driven infrastructure in large-scale industrial environments~\cite{cabane2024impact,lazzari2023event}, as well as the integration of \texttt{SmellHunter} with UML-based software modeling workflows commonly adopted in industry~\cite{junior2022use}. Finally, future versions of the platform will explore machine learning techniques to predict software design problems and recommend proactive refactoring strategies based on contextual smell patterns and historical project data~\cite{silva2023approach}.





\bibliographystyle{elsarticle-num}
\bibliography{references}


%
%
%

\end{document}